\documentstyle[11pt,amstex]{article}                                
\input{amssymb.sty}

\def\AFOUR{%
\setlength{\textheight}{9.0in}%
\setlength{\textwidth}{5.75in}%
\setlength{\topmargin}{-0.375in}%
\hoffset=-.5in%
\renewcommand{\baselinestretch}{1.17}%
\setlength{\parskip}{6pt plus 2pt}%
}

\AFOUR                                           

\makeatletter
\def\section{\@startsection {section}{1}{\z@}{-3.5ex plus -1ex minus
 -.2ex}{2.3ex plus .2ex}{\large\bf}}
\def\subsection{\@startsection{subsection}{2}{\z@}{-3.25ex plus -1ex minus
 -.2ex}{1.5ex plus .2ex}{\normalsize\bf}}
\makeatother

\makeatletter
\@addtoreset{equation}{section}

\makeatother

\newcommand{\nc}{\newcommand}
\newcommand{\rnc}{\renewcommand}

\nc{\bea}{\begin{eqnarray}}
\nc{\eea}{\end{eqnarray}}
\nc{\be}{\bea}
\nc{\ee}{\eea}

\rnc{\a}{\alpha}
\nc{\ab}{\bar{\a}}
\nc{\ap}{\a^{+}}
\nc{\abm}{\ab^{-}}
\rnc{\b}{\beta}
\nc{\bb}{\bar{\b}}
\nc{\bbp}{\bb_{\zb}^{+}}
\nc{\bm}{\b_{z}^{-}}
\nc{\oa}{\overline{\a}}
\nc{\ob}{\overline{\b}}
\rnc{\gg}{\gamma}
\rnc{\d}{\delta}
\nc{\f}{\phi}
\nc{\fb}{\bar{\phi}}
\nc{\vf}{\varphi}
\nc{\p}{\psi}

\rnc{\c}{\chi}
\nc{\la}{\lambda}
\nc{\m}{\mu}
\nc{\n}{\nu}
\rnc{\o}{\omega}
\nc{\Om}{\Omega}
\rnc{\t}{\theta}
\nc{\eps}{\epsilon}
\nc{\F}{\Phi}

\nc{\trac}[2]{{\textstyle\frac{#1}{#2}}}

\nc{\ex}[1]{\mbox{e}^{\,\textstyle#1}}

\nc{\mat}[4]{\left(\begin{array}{cc}#1&#2\\#3&#4\end{array}\right)}

\nc{\som}[9]{\left(\begin{array}{ccc}#1&#2&#3\\#4&#5&#6\\#7&#8&#9%
\end{array}\right)}

\nc{\tr}{\mathop{\mbox{tr}}\nolimits}
\nc{\ad}{\mathop{\mbox{ad}}\nolimits}
\nc{\Tr}{\mathop{\mbox{Tr}}\nolimits}
\nc{\Det}{\mathop{\mbox{Det}}\nolimits}
\nc{\rk}{\mathop{\mbox{rk}}\nolimits}
\nc{\ra}{\rightarrow}
\nc{\Ra}{\Rightarrow}
\nc{\LRa}{\Leftrightarrow}
\nc{\ot}{\otimes}
\rnc{\ss}{\subset}
\nc{\nul}{\noindent\underline}
\nc{\non}{\nonumber\\}

\nc{\subs}[1]{{\vspace*{0.5cm}}%
{\noindent\underline{#1}}{\addcontentsline{toc}{subsection}{#1}}%
{\vspace*{0.3cm}}}

\nc{\zb}{\bar{z}}
\rnc{\lg}{\frak{g}}
\nc{\lt}{\frak{t}}
\nc{\lk}{\frak{k}}
\nc{\lh}{\frak{h}}
\nc{\pik}{\Pi_{\lk}}
\nc{\pip}{\Pi_{+}}
\nc{\pim}{\Pi_{-}}
\nc{\pih}{\Pi_{\lh}}
\nc{\jz}{J_{z}}
\nc{\jzh}{\jz^{\lh}}
\nc{\jzp}{\jz^{+}}
\nc{\jzm}{\jz^{-}}
\nc{\del}{\partial}
\nc{\dz}{\del_{z}}
\nc{\dzb}{\del_{\bar{z}}}
\nc{\az}{A_{z}}
\nc{\azb}{A_{\bar{z}}}
\nc{\g}{g^{-1}}
\nc{\dw}{\Delta_{W}}
\nc{\Ad}{{\mbox{Ad}}}
\nc{\ks}{Ka\-za\-ma-\-Su\-zu\-ki}
\nc{\KS}{\ks}
\nc{\ksm}{\ks\ model}
\rnc{\AA}{{\Bbb A}}
\nc{\BB}{{\Bbb B}}
\nc{\CC}{{\Bbb C}}
\nc{\PP}{{\Bbb P}}
\nc{\cpm}{\CC\PP(m)}
\nc{\cpn}{\CC\PP(n)}
\nc{\cp}[1]{\CC\PP(#1)}
\nc{\gmn}{G(m,m+n)}
\nc{\gmnk}{\gmn_{k}}
\nc{\cO}{{\cal O}}
\nc{\bcO}{\bar{\cO}}
\nc{\bO}{\bar{O}}
\nc{\oQ}{\overline{Q}}
\nc{\ZHS}{{\Bbb Z}{\mathrm HS}}
\nc{\QHS}{{\Bbb Q}{\mathrm HS}}
\nc{\HK}{Hyper-K\"{a}hler\;}
\nc{\RW}{Rozansky-Witten\;}
\nc{\oJ}{\overline{J}}
\nc{\oL}{\overline{L}}
\nc{\oI}{\overline{I}}
\nc{\oK}{\overline{K}}
\nc{\oM}{\overline{M}}
\nc{\oN}{\overline{N}}

\newtheorem{claim}{Claim}[section]

\begin{document}
\global\parskip=4pt

\makeatletter
\begin{titlepage}
\begin{flushright}
IC/2000/9
\end{flushright}
\begin{center}
\vskip .5in
{\LARGE\bf Holomorphic Vector Bundles, Knots and the Rozansky-Witten
Invariants}\\
\vskip 0.4in
{\bf George Thompson}\footnote{email: thompson@ictp.trieste.it}
\vskip .1in
ICTP \\
P.O. Box 586 \\
34100 Trieste \\
Italy\\
\end{center}

\vskip .4in
\begin{abstract}

Link invariants, for 3-manifolds, are defined in the
context of the \RW theory. To each knot in the link one associates
a holomorphic bundle over a holomorphic symplectic manifold $X$. The
invariants are evaluated for $b_{1}(M) \geq
1$ and $X$ Hyper-K\"{a}hler. To obtain invariants of \HK $X$ one finds
that the holomorphic vector bundles must be hyper-holomorphic. This
condition is derived and explained. Some results for $X$ not \HK are presented.

\end{abstract}

\end{titlepage}
\makeatother

\begin{small}
\tableofcontents
\end{small}

\setcounter{footnote}{0}

\section{Introduction}

This paper is concerned with the definition and evaluation of
invariants that can be associated with knots and links in the context
of the \RW model \cite{RW}. This theory has as its basic data a
3-manifold $M$ and a holomorphic symplectic manifold $X$. The path
integral for this theory is a supersymmetric theory based on maps from
$M$ to $X$. I will give a quick review of this in the next
section. For more details of the construction one should consult the
references. 

Rozansky and Witten observed that their theory is a kind of Grassmann odd
version of Chern-Simons theory where, amongst other things, the
structure constants, $f^{a}_{\; bc}$ of the Lie algebra in Chern-Simons
theory go over to $R^{I}_{\; JK\oL}\eta^{\oL}_{0}$ in the \RW model. The
comparisons that are to be made are between the n-th order terms in a
$1/\sqrt{k}$ expansion in Chern-Simons theory and the \RW invariant
evaluated for some dim$_{{\Bbb C}} X = 2n$ \HK
manifold. More precisely, for a $\QHS$ (rational homology sphere), the
n-th order term in the
Chern-Simons theory for group G can be written as
\be
Z^{CS}_{n}[M] = \sum_{\Gamma_{n}} b_{\Gamma_{n}}(G) \, \sum_{a} I_{\Gamma_{n} ,
a}(M) \label{csn}
\ee
while for the ${\mathrm dim}_{{\Bbb C}} X = 2n$ \HK manifold the \RW
invariant reads as
\be
Z^{RW}_{X}[M] = \sum_{\Gamma_{n}} b_{\Gamma_{n}}(X) \, \sum_{a} I_{\Gamma_{n} ,
a}(M). \label{rwn}
\ee
The notation is as follows. The $\Gamma_{n}$ represent all the possible
Feynman graphs of the theory and the sum over the label $a$ is that of
all possible ways of assigning Feynman diagrams to the same graph. The
Feynman diagrams and graphs are the same in the Chern-Simons and \RW
theories. The $I_{\Gamma_{n} ,a}(M)$ are the integrals over $M$ of
products of Greens functions that appear in both theories. 

The interesting part corresponds to the weights $b_{\Gamma_{n}}$ as this is
`all' that distinguishes the two theories. Different weight systems
will yield topological field theories providing the $b_{\Gamma_{n}}$
obey the IHX relations \cite{LMO}. Indeed both the $b_{\Gamma_{n}}(G)$ of
Chern-Simons theory and the $b_{\Gamma_{n}}(X)$ of the \RW theory
satisfy the IHX relations.

While one class of knot observables was defined in
\cite{RW} (and an algorithm given for the associated weights) they
played no essential role there. However, one can show that the
expectation values of Wilson loop observables in the Chern-Simons
theory, and of the knot observables of Rozansky and Witten take a form
analogous to (\ref{csn}) and (\ref{rwn}) respectively. Once more the
differences lie in the weights.  To obtain a topological theory of
knots (and links)
the weights associated with the knot observables need to satisfy the
STU relation. This is clearly satisfied by the Wilson loop observables
in Chern-Simons theory and, as I will show below, also satisfied by
the weights of the knot invariants in the \RW model.

Why write this paper? While they are of interest in
themselves, I believe that, amongst other things, we also need to have
a better understanding of
these observables in order to get at the surgery formulae for the \RW
invariants $Z_{X}^{RW}[M]$ when the \HK manifold $X$ has ${\mathrm
dim}_{{\Bbb C}}X \geq 4$. Of course one of the things one would like
to know about these
invariants is if they are part of the ``universal'' knot invariants
that arise in the LMO construction. Another reason for studying these is that
recently, Hitchin and Sawon \cite{HS} have found
that the \RW theory provides information about \HK
manifolds. Hopefully one will have a bigger set of invariants for
\HK manifolds by allowing for knot observables.

Before going on it is appropriate to ask why should this topological
field theory give us invariants of \HK
manifolds? In order to answer this question let us recall some other
topological field theories.
We know that certain supersymmetric quantum mechanics models yield
information about a manifold $X$ (the models depend on how much extra
structure we are willing to place on $X$). So for example such
supersymmetric models provide simple ``proofs'' of the index
theorems\footnote{In the present setting there are
topological field theories that yield the index formulae for
the Euler characteristic of $X$ (Gauss-Bonnet), the signature of $X$
(Hirzebruch),
if $X$ is a spin manifold for the $\hat{A}$ genus (Atiyah-Singer),
while if $X$ is a
complex manifold one can also obtain the Riemann-Roch formula for the
${\mathrm Todd}$ genus.}. These theories involve maps $S^{1}
\rightarrow X$.  There are also topological field theories which
yield the Gromov-Witten invariants of a complex manifold $X$. These
models are based on holomorphic maps from a Riemann surface to a
compact closed complex $X$, $\Sigma \rightarrow X$.

From this perspective one would expect that a theory based on maps
from a 3-manifold into a \HK $X$, $M \rightarrow X$, would indeed give
rise to invariants for $X$ and that the correct question is, instead,
why do they give invariants of 3-manifolds?

The crux of the matter is that what we learn depends by and large on
what we know. A topological field theory from one point of view is a
theory defined on the space of sections of a certain bundle. To define
the theory it may be necessary to make certain additional choices, for
example, to fix on a preferred Riemannian metric on the total space of
the bundle. In the
best cases this theory will give invariants that do not depend on any
of the particular choices made. What information there is to extract will
depend crucially on the bundle in question. In principle, however, the
topological field theory will provide invariants for the total space
of the bundle. If either the base or the fibre is well understood then
the invariants are really invariants for the fibre or the base
respectively. 

In the case of supersymmetric quantum mechanics we know all there is
to know about $S^{1}$ and so the topological field theories will yield
information about $X$.  The same is true for the Gromov-Witten
theory, since Riemann surfaces are completely classified. The case of
the \RW theory is very different. We do not know much about
3-manifolds nor about \HK manifolds. By fixing on ones favourite \HK
manifold and varying the 3-manifold we get invariants for the
3-manifolds. On the other hand on picking a particular 3-manifold or by
using other knowledge about the 3-manifold invariants and
varying $X$ we learn about $X$.

This paper is organized as follows. In the next section there is a
brief summary of the \RW theory. In section 3 knot and link
observables are introduced. The
expectation values of the link
observables are the link invariants. The concept of a hyper-holomorphic
bundle is seen to arise naturally from the requirement that the
observables will correspond to invariants for \HK $X$. That the
observables do
correspond to link invariants for the 3-manifold $M$ requires that
they satisfy the STU
relation. This relation is derived in section 4. Section 5 is devoted
to stating some explicit results that I have derived, the derivation
being postponed till section 7. Section 6 is by way of a
digression on the properties of the theory when $X=T^{4n}$ while
in section 7 an outline of the proofs is given, the bulk of the work being
defered to the references. Finally, in the appendix, a slightly more
general class of observables is introduced.

All calculations are done using path integrals. The normalization that
I have taken is so that
the \RW for a 3-torus, $T^{3}$ is the Euler characteristic of the \HK
manifold $X$ (or more generally the integral of the Euler density of
$X$ if $X$ is non-compact) \cite{T}.

Some of the results presented here have also been obtained by J. Sawon
\cite{S}.

{\bf Acknowledgments}: Justin Sawon pointed out the relevance of the
work of Verbitsky to
me. I thank him for this and other correspondence. Nigel Hitchin
kindly explained to me the relevance of the \HK quotient
construction as a means for finding hyper-holomorphic bundles. Boris Pioline
brought my attention to the paper \cite{MS} which in turn led me to
Nigel Hitchin. I have benefited
from conversations with M. Blau and L. G\"{o}ttsche. I am grateful to
the referee who suggested numerous improvements. I am especially
indebted to M.S. Narasimhan for his interest and his kind advice at
all stages of this work. This work was supported in part by the EC
under TMR contract ERBF MRX-CT 96-0090.

\section{The \RW Theory}

The construction of the \RW model for holomorphic symplectic $X$ is
described in the appendix of \cite{RW}. I will not need that level of
generality here, though to prove some of the results that I present below
one does need to have the full theory at ones disposal.

The action, for \HK $X$ can be written down without picking a
preferred complex structure from the $S^{2}$ of available complex
structures on $X$. In this way one establishes that the theory yields
invariants of $X$ as a \HK manifold. Since the knot observables,
in any case, require us to make such a choice fix on the complex
structure, $I$, on $X$
so that the $\f^{I}$ are local holomorphic coordinates with respect to
this complex structure. The action is, in the preferred complex structure,
\be
S = \int_{M} L_{1} \sqrt{h} \, d^{3}x\, + \, \int_{M} L_{2}
\label{rwact} 
\ee
where
\bea
L_{1} & = & \frac{1}{2}g_{ij}\partial_{\mu}\f^{i}\partial^{\mu}\f^{j}
+ g_{I\oJ} \chi^{I}_{\mu}D^{\mu} \eta^{\oJ} \label{lag1}\\
L_{2} & = & \frac{1}{2}\left( \eps_{IJ} \chi^{I} D \chi^{J} -
\frac{1}{3} \eps_{IJ}R^{J}_{\; KL\oM} \, \chi^{I}\chi^{K}\chi^{L} \eta^{\oM}
\right). \label{lag2} 
\eea
The covariant derivative is
\be
D_{\mu \, \oJ}^{\oI} = \partial_{\mu}\d^{\oI}_{\oJ} +
(\partial_{\mu}\f^{i})\Gamma^{\oI}_{i \oJ} .
\ee
The tensor $\eps_{IJ}$ is the holomorphic symplectic 2-form that is
available on a \HK manifold. It is a closed, covariantly constant
non-degenerate holomorphic 2-form. Non-degeneracy means that there
exists a holomorphic tensor $\eps^{IJ}$ such that
\be
\eps^{IJ}\, \eps_{JK} = \d^{I}_{K} .
\ee

Both of the Lagrangians $L_{1}$ and $L_{2}$ are invariant under two
independent BRST supersymmetries. In fact $L_{1}$ is BRST exact. These
supersymmetries are also
defined without the need of picking a prefered complex structure on
$X$. But, since a prefered complex structure has already been chosen
in writing the theory, it is easiest to exhibit the BRST operators in
this complex structure. $\overline{Q}$ acts by
\be
\begin{array}{ll}
\overline{Q}\f^{I} = 0, & \overline{Q}\f^{\overline{I}} = \eta^{\oI}, \\
\overline{Q}\eta^{I}= 0, & \overline{Q}\chi^{I} = -d\f^{I} ,
\end{array} \label{brst}
\ee
while $Q$ acts by
\be
\begin{array}{ll}
Q\f^{I} = T^{I}_{\; \oJ}\eta^{\oJ}, & Q\f^{\overline{I}} = 0, \\
Q\eta^{\oI}= 0, & Q\chi^{I} = -T^{I}_{\; \oJ}d\f^{\oJ} -
\Gamma^{I}_{JK}T^{I}_{\; \oJ}\eta^{\oJ}\chi^{K}  ,
\end{array} \label{brst2}
\ee
where $T^{I}_{\; \oJ} = \eps^{IK}g_{K\oJ}$ and represents an
isomorphism between $TX^{(1,0)}$ and $TX^{(0,1)}$. The BRST charges
satisfy the algebra,
\be
\overline{Q}^{2}=0, \;\;\; \{ Q , \overline{Q} \} = 0, \;\;\; Q^{2}=0.
\ee

Set $\eta^{I} = T^{I}_{\; \oJ}\eta^{\oJ}$ in order to make contact with
the notation of the bulk of \cite{RW} and that used in \cite{T} and \cite{HT}.

\section{The Knot Observables}

I will define knot and link invariants by associating holomorphic
bundles over a holomorphic symplectic manifold $X$ to the knot or
link. However, special issues which arise when $X$ is \HK are addressed
in some detail. 

\subsection{Associating the Holomorphic Tangent Bundle}
The observables associated to a knot, ${\mathcal K}$, that were suggested in
\cite{RW} are
\be
{\mathcal O}_{\a}(K) = \Tr_{\a}\, {\mathrm P} \,
\ex{\oint_{K}
A} , \label{kobs}
\ee
where the ${\mathrm sp}(n)$ connection is
\be
A_{\; J}^{I}= d\f^{L}\Gamma^{I}_{LJ} - \eps^{IM}
\Omega_{MJKL}\chi^{K}\eta^{L} ,
\ee
and $\a$ designates a representation $\Upsilon_{\a}$ of ${\mathrm
sp}(n)$. Some properties of the connection are:
\begin{enumerate}
\item The connection is $\overline{Q}$ exact
\be
A_{\; J}^{I}= - \overline{Q} \left( \chi^{L}\Gamma^{I}_{LJ} \right).
\ee
This does not mean that it is a ``trivial'' observable since it is
the BRST variation of a connection.
\item On a general holomorphic symplectic manifold $X$, the connection is
\bea
A_{\; J}^{I} &= & - \overline{Q} \left( \chi^{L}\Gamma^{I}_{LJ}
\right) \nonumber \\
& = & d\f^{L}\Gamma^{I}_{LJ} + R^{I}_{\;JK \overline{L}}\chi^{K}
\eta^{ \overline{L}} 
\eea
where $\Gamma^{I}_{LJ}$ is some symmetric connection on the
holomorphic tangent bundle and
\be
R^{I}_{\;JK \overline{L}} =
\overline{\partial}_{\overline{L}}\Gamma^{I}_{JK} ,
\ee
is the Atiyah class of $X$. The Atiyah class is the obstruction to the
connection being holomorphic \cite{A}.
\item If $X$ is \HK then we might also want to be sure the observable
does not depend on the particular choice of the $S^{2}$ of complex
structures. This is indeed the case since,
\be
Q A_{\; J}^{I} = d_{A} \Lambda^{I}_{\; J} = d \Lambda^{I}_{\; J} +
[ A, \Lambda ]^{I}_{\; J} ,
\ee
where $\Lambda^{I}_{\; J} = \eta^{L} \Gamma^{I}_{LJ}$. A $Q$
transformation is, therefore, equivalent to a gauge transformation and we are
assured that the Wilson loop is $Q$ invariant since it is gauge
invariant. (The situation will be made clearer below)

\end{enumerate}

\subsection{Associating Holomorphic Vector Bundles}

Let $E \rightarrow X$ be a holomorphic vector bundle over a holomorphic
symplectic $X$ with fibre $V$. The reason for choosing $E$ to be a
holomorphic bundle is that we want the STU relations to be satisfied,
see section \ref{STU}. In trying to mimic the construction of
the observables
(\ref{kobs}) we will find some more stringent conditions on $E$. Let $\o$
be a connection on $E$ whose, on fixing the complex structure of $X$,
$(0,1)$ component, in a holomorphic frame, vanishes that is 
\bea
\o &=& \o^{(1,0)} , \nonumber \\
&=& \o_{I} \, dz^{I} .
\eea
Since $\Phi : M \rightarrow X$ one can pull $E$ back to the 3-manifold
$M$ and consider the connection
\bea
A & = & - \overline{Q} \left( \chi^{I} \o_{I} \right) \nonumber \\
 &=& d\f^{I}\o_{I} +  \overline{\partial}_{\overline{J}}\o_{I} \, \chi^{I}
\eta^{ \overline{J}} , \label{holcon}
\eea
which, due to the presence of the fermion terms, is not just the usual pull back $\Phi^{*}(\o)$. Associate to a
knot the observable
\be
{\mathcal O}_{E}(K) = \Tr_{V}\, {\mathrm P} \,
\exp{\left(\oint_{K} A\right)} . \label{hobs}
\ee
Next we list the
relevant properties of this connection and specify extra requirements
on $E$ so that we obtain a good observable:
\begin{enumerate}
\item The connection $A$ is $\overline{Q}$ exact but non-trivial.
\item If $X$ is \HK and one wants (\ref{hobs}) to also be invariant under $Q$
(and so not to depend on the choice of complex structure on $X$) then
the connection must satisfy
\be
F^{(2,0)}_{\o} = 0 , \label{hermitian}
\ee
as well as
\be
T^{\overline{J}}_{K}
\, \overline{\partial}_{\overline{J}}\,  \o_{I} = T^{\overline{J}}_{I}
\, \overline{\partial}_{\overline{J}} \, \o_{K} . \label{hyphol}
\ee
The condition (\ref{hermitian}) can be satisfied by choosing a
Hermitian metric on the bundle and then taking $\o$ to be the unique
hermitian connection.  
Holomorphic bundles $E$ that also satisfy (\ref{hyphol}) are said to
be 
hyper-holomorphic: a term coined by Verbitsky \cite{V}. It is an immediate
consequence that such bundles are {\bf stable}, since contracting
(\ref{hyphol}) with $\eps^{IK}$ yields
\be
g^{I \overline{J}} F_{I \overline{J}} = 0 , \label{stable}
\ee
which was conjectured to be equivalent to the condition of stability by
Hitchin and Kobayashi and proven to be so by Donaldson and Uhlenbeck
and Yau.
One easy result, by counting equations, is that (\ref{hyphol}) and
(\ref{stable}) are equivalent when $X$ is a \HK surface. There is an
important converse due to Verbitsky \cite{V}. Let $E$ be a stable
holomorphic bundle over a \HK 
manifold for a given complex structure $I$ then, if $c_{1}(E)$ and
$c_{2}(E)$ are invariant under the natural $Sp(1)$ action, $E$ is
hyper-holomorphic.

If the holomorphic bundle is hyper-holomorphic then
\be
QA = d_{A} \Lambda , \;\;\;\; \Lambda = \eta^{\oI} \, T^{\, J}_{\oI}
\, \o_{J} , \label{gt}
\ee
and, consequently, a $Q$ transformation is equivalent to a gauge
transformation. 
\end{enumerate}

\noindent{\bf Remark}: $Q$ and $\oQ$ are the components of the
${\mathrm sp}(1)$ doublet BRST operator $Q_{A}$ in the preferred complex
structure for $X$ (For this see equations (2.17) to (2.22) in
\cite{RW}.) Invariance under both of these operators means
invariance under the action of $Q_{A}$. However, $Q$ and $\oQ$ are
essentially to be identified with the twisted Dolbeault operators,
$\partial_{\o}$ and $\overline{\partial}$ respectively, in the preferred
complex structure. Invariance under $Q_{A}$ means that if we choose a
different complex structure (say $I'$) then the knot observables will
be invariant under the corresponding BRST operators $Q'$ and $\oQ'$,
which in turn are to be identified with $\partial_{\o}'$ and
$\overline{\partial}'$.

\noindent{\bf Remark}: If one is only interested in obtaining
3-manifold invariants, then all one really requires is that $E$ be
holomorphic with respect to the given complex structure $I$ on $X$.

\subsection{The Meaning of (\ref{hyphol}) and Hyper-holomorphic Vector
Bundles}
There is a nice geometric interpretation of the equation
(\ref{hyphol}), already mentioned above, which is that it is the condition
for which the holomorphic vector bundle $E$ is holomorphic for
the entire sphere's worth of complex structures. Let us see that this is
the case in a more mundane manner. 

Let $I$ be a given complex structure on $X$. The
complexified tangent and cotangent bundles, $TX_{{\Bbb C}}$ and $T^{*}
X_{{\Bbb C}}$ split into a sum of holomorphic and anti-holomorphic
bundles as $T^{(1,0)}X \oplus T^{(0,1)}X$ and $T^{*(1,0)}X \oplus
T^{*(0,1)}X$ respectively. The decomposition is such that $(1-iI)/2 : T^{*}
X_{{\Bbb C}} \rightarrow T^{*(1,0)}X$.

Since $X$ is \HK, with complex structures
$I$, $J$, and $K$ satisfying the usual quaternionic rules,  $I'=I + \d I
= I +  \d b J + \d c K$ is an infinitesimally deformed complex
structure. Denote the the splitting of $T^{*}
X_{{\Bbb C}}$ with respect to $I'$ by $T^{*(1,0)}X' \oplus
T^{*(0,1)}X'$. If $dz$ is a basis for $T^{*(1,0)}X$,  $d\overline{z}$
is a basis for $T^{*(0,1)}X$, $dw$ is a
basis for $T^{*(1,0)}X'$ and $d\overline{w}$ is a basis for
$T^{(0,1)}X'$ we find
\be
dw = \left( 1 - \alpha \overline{T} \right) \, dz + \overline{\alpha} \, T \,
d\overline{z} 
\ee
where $\alpha = (\d c + i \d b)/4)$ and $T = (J-iK):T^{*(0,1)}X \rightarrow
T^{*(1,0)}X$ or put another way $Td\overline{z} \in {\mathrm H}^{1}(X,
T^{(1,0)}X)$. All of this fits within the Kodaira-Spencer theory of
complex deformations. The spheres worth of complex structures means
that we only look at the one (complex) dimensional subspace of
${\mathrm H}^{1}(X, T^{(1,0)}X)$ spanned by $T$.

Let us now pass on to the case of holomorphic vector bundles. Our
holomorphic bundles over $X$ come equipped with a connection $\o$ that
satisfies
\bea
& & \o^{(0,1)} = 0 \\
& & F_{\o}^{(2,0)}=0,
\eea
where the holomorphic splitting is with respect to the given complex
structure, I, on $X$. 

In this section by a hyper-holomorphic vector bundle I will
mean a holomorphic vector bundle equipped with a given connection
which has curvature of type $(1,1)_{{\mathcal J}}$ for {\bf all} of the
${\mathcal J} \in S^{2}$ of complex structures
on $X$. Now suppose that we want that $E$ be
hyper-holomorphic this means, in particular, that for the deformed
complex structure $I'$, that $F_{\o}$ be of type $(1,1)'$. Given that
$F_{\o}$ with respect to $I'$ should be of type $(1,1)'$ but of type
$(1,1)$ with respect to $I$ means that one gets
conditions on the $(1,1)$ component of the curvature. These conditions
are obtained on perusal of the following
\bea
F_{\o}^{(1,1)'} 
& =&  F_{I \overline{J}}(\o)\, dw^{I} \,
d\overline{w}^{\overline{J}} \nonumber \\
& =&  F_{I \overline{J}}(\o) \left( dz -\alpha \overline{T} \,
dz +  \overline{\alpha}\, T\, d\overline{z} \right)^{I}\, \left
( d\overline{z}  -
 \overline{\alpha}\, T\, d\overline{z} + 
\alpha \overline{T} \,
dz  \right)^{\overline{J}} \nonumber \\
& =&  F_{I \overline{J}}(\o) \,\left(  dz^{I} \,
d\overline{z}^{\overline{J}} - dz^{I} \overline{\alpha} \,( T\,
d\overline{z})^{\overline{J}}+  dz^{I} \alpha(\overline{T} \,
dz)^{\overline{J}}\right. \nonumber \\
& & \;\;\;\;\;\;\; \;\;\;\;\;\;\;\; \;\;\;\;\;\;  \left. - \alpha
(\overline{T}  \,
dz)^{I} d\overline{z}^{\overline{J}}  +   \overline{\alpha}(T
d\overline{z})^{I}d\overline{z}^{\overline{J}} \right) ,
\eea
the $(0,2)$ and $(2,0)$ components on the right hand side will vanish iff
\be
T^{I}_{\overline{K}}F_{I\overline{J}}(\o)
=T^{I}_{\overline{J}}F_{I\overline{K}}(\o) . \label{hyphol1}
\ee
These are precisely the equations (\ref{hyphol}) that we found in the
previous section and so the current definition of a hyper-holomorphic
vector bundle agrees with that of Verbitsky given in the previous section.


What is particularly satisfying is that the physics, demanding that
(\ref{hobs}) also be invariant under $Q$, leads naturally to this
definition of a hyper-holomorphic bundle. This is the way I
came to it before I was informed that this definition had already
appeared in the mathematics literature \cite{V}. Indeed the definition
is older than this reference having already appeared in \cite{MS} and
the demonstration that the bundles in question are
hyper-holomorphic is attributed, in that reference, to N. Hitchin.

\subsection{On The Existence of Hyper-Holomorphic Bundles}

Clearly the holomorphic tangent bundle of a \HK manifold is
hyper-holomorphic, but apart from on a \HK surface I did not know of any
general results on the existence of hyper-holomorphic
bundles. N. Hitchin \cite{H} has kindly answered the following question in the
affirmative: Are there examples of hyper-holomorphic
bundles over a hyper-K\"{a}hler $X$ other than its holomorphic tangent
bundle? Indeed he shows that there is a procedure for
constructing such bundles which follows directly from the
hyper-K\"{a}hler quotient construction \cite{HKLR}. The details of the
construction have also appeared in \cite{GN}. I give a very brief
description of the salient features.

Let $G$ be a compact Lie group acting on a hyper-K\"{a}hler manifold
$Y$, with either $H^{1}(Y, {\Bbb Z})=0$, or $H^{2}(G)=0$,
which preserves both the metric and the hyper-K\"{a}hler
structure. Consequently the group preserves the three K\"{a}hler
forms, $\o_{A}$, corresponding to the three complex structures $A =
I$, $J$, $K$. For each K\"{a}hler
form there is an associated moment map, $\mu_{A}: \rightarrow
\lg^{*}$, to the dual vector space $\lg^{*}$ of the Lie algebra.

Each element $\zeta$ of the Lie algebra $\lg$ of $G$
defines a vector field, denoted $\overline{\zeta}$, which generates the
action of $\zeta$ on $Y$. Then, up to a constant for connected $Y$,
\be
d\mu^{\overline{\zeta}}_{A} = i_{\overline{\zeta}}\o_{A},
\ee
defines $\mu^{\overline{\zeta}}_{A}$. The moment maps $\mu_{A}$ are defined by
\be
\langle \mu_{A}(m), \zeta \rangle = \mu^{\overline{\zeta}}_{A}(m) ,
\ee
and they can be grouped together into one moment map
\be
\mu : Y \rightarrow {\mathbf R}^{3} \otimes \lg^{*} .
\ee

{\bf Fact 1} \cite{HKLR}: For any $\zeta^{*}\in {\mathbf R}^{3}\otimes
\lg^{*}$ fixed by the action of $G$, the quotient space $X =
\mu^{-1}(\zeta^{*})/G$ has a natural Riemannian metric and
hyper-K\"{a}hler structure.

{\bf Fact 2} \cite{H, HKLR, GN}: Let $\pi:
\mu^{-1}(\zeta^{*}) \rightarrow \mu^{-1}(\zeta^{*})/G = X$ be the
projection. Then, $\pi: \mu^{-1}(\zeta^{*}) \rightarrow X$ is a
principal G-bundle which comes equipped with a natural connection
$\Theta$, where the horizontal space is the orthogonal complement of
the tangent space of the orbit $T_{y}\mu^{-1}(\zeta^{*})$ with $y \in 
\mu^{-1}(\zeta^{*})$.

{\bf Fact 3} \cite{H, GN}: The natural connection is
hyper-holomorphic.

The upshot is that if the hyper-K\"{a}hler manifold of interest comes
from a hyper-K\"{a}hler quotient construction then it comes equipped
with a natural hyper-holomorphic principal bundle. Given a
representation of $G$ we can construct an associated hyper-holomorphic
vector bundle, which is what we are after. In the case of infinite
dimensional quotients (on the space of connections for example) the
associated hyper-holomorphic vector bundles are index bundles,
universal bundles, etc.. Explicit examples in the case of the monopole moduli
space can be found in \cite{MS} while for instantons one may refer to
\cite{GN}.

\section{The STU Relation}\label{STU}
In the physics approach to topological field theory it is formally
enough that one can exhibit metric independence via standard physics
arguments. (A metric variation is BRST exact, for example, which is
the case in Chern-Simons theory when one includes gauge fixing terms.) The
essence of the argument in the case of Chern-Simons theory for
3-manifold invariants has been distilled, made mathematically precise
and then abstracted. The net result is the so called IHX
relation. 

A crucial feature of the \RW theory is that the IHX relation is
satisfied by the weights $b_{\Gamma}(X)$. A proof of this statement
for $\QHS$ goes along the following lines (this is taken from
\cite{RW}). Vertices in a closed $2n$-vertex graph in this theory
carry the curvature tensors $R^{I}_{\; JK\oL}$. Their holomorphic
labels are contracted with $\eps^{IJ}$ (thanks to the $\chi$
propagator). The anti-holomorphic labels are totally anti-symmetrized
(since this involves products of $\eta^{\oI}_{0}$) and from the
Bianchi identity one has $\overline{\partial}_{\oM}R^{I}_{\; JK\oL}=
\overline{\partial}_{\oL}R^{I}_{\; JK\oM} $ and so one obtains a
$\overline{\partial}$-closed $(0,2n)$ form on $X$, that is a map
\be
\Gamma_{n,3} \rightarrow {\mathrm H}^{2n}(X) .
\ee
The weight functions $b_{\Gamma}(X)$ satisfy the IHX relations by
virtue of the fact that
\be
\overline{\partial}_{\oN} \nabla_{M}R^{I}_{\; JK\oL}
\eta_{0}^{\oN}\eta_{0}^{\oL} = \left(R^{I}_{\; PM\oN}R^{P}_{\;
JK\oL}+R^{I}_{\; PK\oN}R^{P}_{\; JM\oL}+ R^{I}_{\; JP\oN}R^{P}_{\;
KM\oL}\right)\eta_{0}^{\oN}\eta_{0}^{\oL} \label{ihx}
\ee
This tells us that the right hand side is cohomologous to zero.

Chern-Simons theory has the IHX relation, essentially the
Jacobi identity for the Lie algebra used in the definition of the
theory, encoded in it in two different ways. Firstly, it is subsumed in
the whole construction of gauge theories. Secondly, it is explicitly
required in order for the BRST operator, $Q$, to be nilpotent
$Q^{2}=0$. How is the IHX relation ``built in'' in the \RW theory? More
concretely how is (\ref{ihx}) manifest from the beginning? It is clear
from (\ref{brst}) that it is not required for nilpotency of the operator
$\overline{Q}$. However, the IHX relation (\ref{ihx}) follows from the
Bianchi identity for the curvature form and it is the Bianchi identity
that ensures that the action (\ref{rwact}) is BRST invariant. So in
this sense the IHX relation is subsumed from the start. It is in this
way that the formal physics proofs of metric independence and the
mathematical proofs are connected.

Now in order to have a good knot or link invariant one would like the
analogue of the STU relation (see for example \cite{B}). In
Chern-Simons theory this amounts to the Lie algebra commutation
rules. Let $T_{a}$ be a basis of
generators for the Lie algebra in the ${\mathbf T}$
representation. Basically the STU relations says, $[T_{a}, T_{b}] =
f^{c}_{ab} T_{c}$. The representation matrices are attached to the
3-point vertices in the loop observables. In the present context the
STU relation is the following,
\be
\overline{\partial}_{\oK} \nabla_{L}F_{\; I\oJ}
\, \eta_{0}^{\oJ}\eta_{0}^{\oK} = -\left( R^{N}_{\; IL\oK}\, F_{
N\oJ}+ [F_{L\oK} ,\, F_{I\oJ}] \right)\eta_{0}^{\oJ}\eta_{0}^{\oK},
\label{stu} 
\ee
again this equation tells us that the right hand side is cohomologous
to zero. In analogy to the Chern-Simons theory the curvature tensor
plays the role of the structure constants and the curvature 2-form the
role of the representation matrices. 

One derives (\ref{stu}) from the
Bianchi identity for the curvature two form of the holomorphic vector
bundle, as follows: Let $E$ be a holomorphic vector bundle, choose
the connection so that $\overline{\partial}_{\o}
=\overline{\partial}$, then\footnote{If we choose a Hermitian
structure we could then fix on the unique Hermitian connection for
which $\overline{\partial}_{\o} =\overline{\partial}$ and
$F_{\o}^{(2,0)}= F_{\o}^{(0,2)}=0$.} 
\be
F_{\o}^{(0,2)} = 0.
\ee
The Bianchi identity $d_{\o}F_{\o}=0$ tells us that
\be
\partial_{\o} F_{\o}^{(2,0)} =0, \;\;\; \partial_{\o} F_{\o}^{(1,1)} =
\overline{\partial} F_{\o}^{(2,0)}, \;\;\; \overline{\partial}
F_{\o}^{(1,1)} = 0 . \label{bianc}
\ee
We want to get a formula for
\be
\overline{\partial}_{\oK} \nabla_{L}(\o) F_{I \oJ} - ( \oK
\leftrightarrow \oJ ) = [\overline{\partial}_{\oK}, \nabla_{L}(\o)]\,
F_{I  \oJ} + \nabla_{L}(\o)\overline{\partial}_{\oK} F_{I \oJ} - ( \oK
\leftrightarrow \oJ ) .
\ee
The last term in this equation vanishes by virtue of the last equality
in (\ref{bianc}), so that
\bea
\overline{\partial}_{\oK} \nabla_{L}(\o) F_{I \oJ} - ( \oK
\leftrightarrow \oJ ) & = & [\overline{\partial}_{\oK}, \nabla_{L}(\o)]\,
F_{I  \oJ}  - ( \oK \leftrightarrow \oJ ) \nonumber \\
& = & - R^{N}_{I L \oK} \, F_{N\oJ} - [F_{L \oK}\, , \, F_{I \oJ} ] - ( \oK
\leftrightarrow \oJ ) ,
\eea
as required.

\section{Claims}

We are interested in evaluating, for a knot $K$ and holomorphic vector
bundle $E$,
\be
Z_{X}[M, \, {\mathcal O}_{E}(K)] = \int D\Phi \, \ex{-S(\Phi)} \,
{\mathcal O}_{E}(K) , \label{wloop}
\ee
and, more generally, for a link made up of a union of non-intersecting
knots $K_{i}$ with a holomorphic vector bundle $E_{i}$ associated to
each knot
\be
Z_{X}[M, \, \prod_{i}{\mathcal O}_{E_{i}}(K_{i})] = \int D\Phi \,
\ex{-S(\Phi)}  \,\prod_{i} {\mathcal O}_{E_{i}}(K_{i}) . \label{hwloop}
\ee
The linking number, ${\mathrm Link}(K_{i},K_{j})$, of the knots $K_{i}$ and $K_{j}$
in a $\QHS$ 
makes an appearance and I use a definition tailored to our present
needs. Let $K_{i}$ denote the $i-$th knot in a ${\Bbb Q}$HS $M$. Since
$H_{1}(M, {\Bbb Q})=0$, we have that
${\rm H}_{1}(M, {\Bbb Z})$ is a finite group and the integral homology classes
represented by the $K_{i}$ are of finite order, say of order $m_{i}$,
so that $m_{i}K_{i}$ (no sum over $i$) is null-homologous. Let $m_{i}J_{i}$
be the de Rham 2-currents Poincare dual to the $m_{i}K_{i}$. Then we have
that $m_{i}J_{i}$ is trivial and so there exist
$\mu_{i}$ such that 
\be
d\mu_{i} = m_{i}J_{i} .
\ee
Observe that the singular support of $m_{i}J_{i}$ (resp. $\mu_{i}$) does
not intersect the singular support of $d\mu_{k}$ (resp. $dJ_{k} =0$)
for $i \neq k$  (see
\cite{dR} $\S 20$ (e) for details).
Set $\la_{i} = \mu_{i}/m_{i}$. The linking number is now defined to be
\be
{\mathrm Link}(K_{i},K_{j}) = \int_{M} \la_{i}\, J_{j} =  \int_{M}
\la_{j}\,  J_{i} = {\mathrm Link}(K_{j},K_{i}). \label{linking}
\ee

In section \ref{calcs} I will prove some of the following claims. $M$
is a 3-manifold and, for the first three claims, 
$X$ is a \HK manifold and the $E_{i}$ are holomorphic vector bundles over $X$
associated to a link.
\begin{claim}\label{cl1}If $b_{1}(M)
\geq 2$ then
\be
Z^{RW}_{X}[M, 
\prod_{i}{\mathcal O}_{E_{i}}(K_{i}) ] = \a \, .\,
Z^{RW}_{X}[M],
\ee
with
\be
\a = \prod_{i}{\mathrm rank(E_{i})}.
\ee
\end{claim}
We need some more notation. Let $F_{\omega_{i}}$ denote the curvature
2-form of $E_{i}$,
\be
{\mathrm ch}(E_{i},t) = {\mathrm Tr}_{V_{i}} \,
\ex{\frac{tF_{\omega_{i}}}{2\pi \sqrt{-1}}} .
\ee
For a holomorphic line bundle $L$ and $t \in {\Bbb Z}$ one has
${\mathrm ch}(L,t) = {\mathrm ch}(L^{\otimes t})$.
   
When $X$ is \HK denote the Chern roots of the holomorphic tangent
bundle by $\pm x_{1}, \dots , \pm x_{n}$. Denote by $\Delta_{M}(t)$
the Alexander polynomial of $M$ normalized so as to be symmetric in
$t$ and $t^{-1}$ and so that $\Delta_{M}(1)= | {\mathrm Tor}H_{1}(M,
{\Bbb Z})|$ and set
\be
 \Delta_{M}(X) = \prod_{i=1}^{n} \Delta_{M} \left( \ex{x_{i}} \right).
\ee

\begin{claim}\label{cl2}If $b_{1}(M)=1$ then 
\be
Z^{RW}_{X}[M, 
\prod_{i}{\mathcal O}_{E_{i}}(K_{i}) ] = -\int_{X} \hat{A}(X) \,
 \Delta_{M}(X) \, \, \prod_{i}{\mathrm
ch}\left(E_{i}, \omega(K_{i})\right) ,
\ee
where $\omega$ is the generator of ${\mathrm H}_{1}(M, {\Bbb Z})$ and
\be
\omega(K_{i}) = \int_{K_{i}} \omega ,
\ee
is the intersection of the Poincare dual of the knot with
$\omega$.
\end{claim}
\begin{claim}\label{cl3}If $M$ is a $\QHS$ and ${\mathrm dim}_{{\Bbb
C}}X =2$ we have that
\bea
Z^{RW}_{X}[M,\prod_{i}{\mathcal O}_{E_{i}}(K_{i}) ] &=&
\a \, .\, Z^{RW}_{X}[M] + |{\mathrm H}_{1}(M, {\Bbb Z})| \left( 
\frac{1}{2}\sum_{i}\a_{i}\, {\mathrm Link}(K_{i}, K_{i})\int_{X}
{\mathrm ch}(E_{i}) \right.
\nonumber \\
& & \;\;\;\;\;\;\; \left. +
\sum_{i<j} \a_{ij}\, {\mathrm Link}(K_{i},
K_{j}) \int_{X} c_{1}(E_{i})c_{1}(E_{j}) \right) \,  ,
\eea
where
\be
\a_{j} = \prod_{k\neq j} {\mathrm rank}(E_{k}), \,\,\,\, \a_{ij} =
\prod_{k\neq i,j} {\mathrm rank}(E_{k}) .
\ee
\end{claim}
{\bf Remark:}
For $S^{2}\times S^{1}$, a knot can wrap say $k$ times around the
$S^{1}$, so that $\omega(K_{i})=k_{i}$ and we have
\be
Z^{RW}_{X}[S^{2}\times
S^{1},\prod_{i}{\mathcal O}_{E_{i}}(K_{i}) ]  = -\int_{X} \hat{A}(X)
\,  \prod_{i=1}{\mathrm
ch}\left(E_{i}, k_{i}\right) . \label{RR}
\ee
This partition function with knot observables can be understood, for
compact $X$ and $k_{i} =1$, as the index of the twisted Dolbeault
operator, coupled to $\prod_{i}\otimes E_{i}$ (see the
appendix). In fact the
Rozansky-Witten path integral yields a proof of the Riemann-Roch
formula for the index of the twisted Dolbeault operator.

There are two more claims that I will not prove, but that can be
established by slight variations of the proofs for the claims above. In
these claims $X$
is a holomorphic symplectic manifold, $M$ a 3-manifold and the $E_{i}$
are holomorphic vector bundles over $X$ associated to a link. The \HK
condition on $X$ is dropped.
\begin{claim}\label{cl4}For $b_{1}(M)>3$
\be
Z^{RW}_{X}[M] = 0.
\ee
\end{claim}
\begin{claim}\label{cl5}
If $b_{1}(M)
\geq 2$ then $Z^{RW}_{X}[M, 
\prod_{i}{\mathcal O}_{E_{i}}(K_{i}) ] = \left(\prod_{i}{\mathrm rank(E_{i})}
\right) \, .\,
Z^{RW}_{X}[M]$.
\end{claim}

{\bf Remark:}
Once more we see that these invariants, for $b_{1}(M)>0$, are
essentially classical invariants of the 3-manifold. To get something
new one must take $M$ to be a $\QHS$.

\section{Some Observations on the Invariants for X a 4n-Torus}

At first sight it is quite odd to realize that while the \RW
invariants vanish for any 4n-torus (since the
curvature tensor vanishes) this is not true for the link invariants. A
glance at claim {\bf \ref{cl2}} shows us that instead, providing $b_{1}(M)
\leq 1$, that $Z^{RW}_{T^{4n}}[M, \prod_{i}{\mathcal O}_{E_{i}}(K_{i})
]$ need not vanish. Indeed for $b_{1}(M) =1$ we have
\be
Z^{RW}_{T^{4n}}[M, 
\prod_{i}{\mathcal O}_{E_{i}}(K_{i}) ] = -\int_{X} \prod_{i}
 \, {\mathrm
ch}\left(E_{i}, \omega(K_{i})\right) ,
\ee
though the right hand side of this expression has very little dependence
on the 3-manifold $M$. We do not fare much better with $M$ a $\QHS$
either as we see next.

\subsection{The \RW Path Integral For X a 4n-Torus and M a $\QHS$}

For $T^{4n}$ the path integral can be exactly performed since the
theory is a ``Gaussian''. Fix on the standard flat metric on
$T^{4n}$. With this choice the metric connection on the holomorphic
tangent bundle and the corresponding Riemann curvature tensor vanish.
Consequently, the path integral becomes
\be
Z_{T^{4n}}[M, \, \prod_{i}{\mathcal O}_{E_{i}}(K_{i})] = \int D\Phi \,
\ex{-S_{0}(\Phi)} \,\prod_{i}{\mathcal O}_{E_{i}}(K_{i}) ,
\ee
where
\be
S_{0}= \int_{M} \left(\frac{1}{2}\d_{ij}\, d\f^{i} * d\f^{j} +
\eps_{IJ} \chi^{I} * d\eta^{J} + \frac{1}{2} \eps_{IJ} \chi^{I}d
\chi^{J}     \right) .
\ee
The fields that appear in the link observable are the the field
$\chi^{I}$ the constant map $\f^{i}_{0}$ and the constant
$\eta^{\oJ}_{0}$. 

The STU relation (\ref{stu}) for tori reads (in Dolbeault cohomology)
\be
[ F_{I\oJ} , F_{J\oK}]\, \eta^{\oJ}_{0}\eta^{\oK}_{0} \sim 0 ,
\ee
which means that the matrices (irrespective of the holomorphic label),
when evaluated in the path integral,
are essentially commuting. Consequently one can drop the path ordering
and simply use the exponential
\be
\Tr_{V_{i}}\, {\mathrm P}\, \exp{\left( i\oint_{K_{i}} A\right)} \sim
\Tr_{V_{i}}\, \exp{\left( i\oint_{K_{i}} A\right)} .
\ee
In order to proceed I use a standard `trick'. Write
\be
\Tr_{V}\, \exp{\left(i\oint_{K} A\right)} = \sum_{D}\, \langle C_{D} |
\exp{\left( i
\oint_{K} A^{A}_{\; B} \,  \overline{C}^{B}  \, C_{A} \right)} \, |
\overline{C}^{D} \rangle , \label{trick}
\ee
where $C_{A}$ and $\overline{C}^{A}$ are Grassmann odd operators with values
in $V$ and $V^{*}$ (the dual vector space) respectively. The operators
$C_{A}$ and $\overline{C}^{A}$ satisfy the usual algebra
\be
\left\{ C_{A} , \overline{C}^{B} \right\} = \d^{B}_{A} ,
\ee
and the states are defined by
\be
C_{A} | 0 \rangle = 0, \;\;\;\;\; {\mathrm and}\;\;\; \overline{C}^{B}
| 0 \rangle = | \overline{C}^{B}\rangle .
\ee
Introduce such variables for each knot $K_{i}$ and index them also
with the label $i$. Then the effective action in the path integral is
\be
S = S_{0} + i\sum_{i} \int_{M} \overline{C}_{i}\, A_{i}\, C_{i}\, J_{i} .
\ee

We can perform the path integral to obtain, up to an integration over
the zero modes,
\be
\sum_{D}\, \langle C_{D} |\exp{\left(-\frac{1}{2} \sum_{i,j} {\mathrm
Link}(K_{i},K_{j})\,  \eps^{IJ}\,  (A_{Ji}\overline{C}_{i}C_{i}) \, 
(A_{Ij}\overline{C}_{j}C_{j}) \right) }\,  |\overline{C}^{D} \rangle
, \label{lob}
\ee
where,
\be
A_{I\, j} =\eta^{\oK} \, \overline{\partial}_{\oK}\,  \o_{I\, j},
\ee
with
\be
\o_{j},
\ee
the connection on the bundle $E_{j}$. A quick way to arrive at this
formula is to use the equation of motion
\be
d\chi^{I} = i \eps^{IJ} \sum_{i} A_{J\, i} \, J_{i} \,
\overline{C}_{i} \, C_{i} ,
\ee
the fact that the equation of motion saturates a Gaussian integral
and to recall (\ref{linking}).

Since the self linking number appears in the formula (\ref{lob}) one
must fix on
some framing of the knots that form the link. Preliminary calculations
\cite{HT2} indicate that the theory comes prepared with the framing
for which the self linking numbers are zero. If one takes, for
simplicity, the framing for which ${\mathrm Link}(K_{i},K_{i}) = 0$,
(\ref{lob}) becomes
\be
{\mathrm Tr}_{\otimes_{i} V_{i}} \, \exp{\left(- \sum_{i<j} {\mathrm
Link}(K_{i},K_{j})\,  \eps^{IJ}\, A_{Ji} \otimes A_{Ij}  \right )} ,
\ee
in general one has 
\be
{\mathrm Tr}_{\otimes_{i} V_{i}} \, \exp{\left(-
\frac{1}{2}\sum_{i j}
{\mathrm Link}(K_{i},K_{j})\,  \eps^{IJ}\, A_{Ji} \otimes A_{Ij}
\right )}, \label{rwlink} 
\ee
with some framing understood. Only the n-th term in the expansion of the
exponential will survive the $\eta^{\oI}_{0}$ integration. 

For example, consider a single knot with self-linking
number $L$, then (\ref{rwlink}), after integration over all the modes,
becomes
\be
|{\mathrm H}_{1}(M, {\Bbb Z})|^{n} L^{n} \int_{X}\, {\mathrm ch}(E),
\ee
so that the invariant enters in a trivial way, meaning that it is
not so interesting as an invariant for \HK manifolds. Recall that,
the \RW invariants for a rank zero 3-manifold (i.e. $b_{1}(M)=0$) do
not depend on $X$
simply through its Chern numbers. If they did there would be precious
few invariants. In the example that we have just considered we have
seen that for a knot and $X$ a torus the holomorphic bundle enters
only through its Chern numbers.

\subsection{Comparing with Chern-Simons Theory}

Now consider the $U(1)$ Chern-Simons theory. There is no perturbation
expansion beyond the lowest order (the theory is quadratic). The
lowest order term is essentially the square root of the inverse of the
Ray-Singer Torsion of $M$. On the \RW side the compact manifolds for
which the \RW invariant vanishes are clearly 4-tori, since the
curvature tensor vanishes, and products of
compact \HK manifolds with a 4-torus. From this point of view the
lowest order in perturbation theory is the 0-torus (point), but this
is hardly insightful. 

It is possible to compare not just the ``pure'' theories but also
those with knot or link observables as well. For Chern-Simons theory
one can introduce Wilson loops
\be
\prod_{j}\exp{\left(iq_{j} \oint_{K_{j}} A\right)} ,
\ee
where the $q_{j}$ are charges. The path integral can again be
evaluated directly and yields, up to a factor of the Ray-Singer
torsion, 
\be
 \exp{\left(-i\frac{2\pi}{k}\sum_{j} q_{j}^{2}\, {\mathrm
L}(K_{j},  K_{j})- \frac{4\pi}{k}\sum_{i < j} q_{i} q_{j}\,  {\mathrm L}(K_{i},
K_{j})\right) }
. \label{cslink}
\ee
As before some framing must be chosen for the self linking numbers
${\mathrm L}(K_{j}, K_{j})$. Expand the exponential
(\ref{cslink}) out to n-th order. Let products of the linking numbers
be a ``basis'' for which to group the terms that arise in such an
expansion. The coefficients will be certain polynomials in the
charges. The comparison with (\ref{rwlink}) can now be completed, the
only difference is in the coefficients, here they are polynomials in
the charges while in the \RW theory they are integrals of products of
curvature 2-forms.

As an example let $n=1$. Then on the Chern-Simons front we get
\be
- i\frac{2\pi}{k}\sum_{j} q_{j}^{2}\, \centerdot \,  {\mathrm
L}(K_{j},  K_{j})- i\frac{4\pi}{k}\sum_{i < j} q_{i} q_{j} \,
\centerdot\,  {\mathrm
L}(K_{i},  K_{j}), 
\ee
while on the \RW side we have, up to a factor of the first homology
group of $M$,
\be
\frac{1}{2}\sum_{j} \a_{j}\, \int_{X}{\mathrm ch}(E_{j})\,
\centerdot\,  {\mathrm
L}(K_{j},  K_{j})- \sum_{i < j} \a_{ij}\,\int_{X}c_{1}(E_{i})\,  c_{1}(E_{j})
\,  \centerdot\, {\mathrm L}(K_{i}, K_{j}), \label{rw1}
\ee
where
\be
\a_{j} = \prod_{k\neq j} {\mathrm rank}(E_{k}), \,\,\,\, \a_{ij} =
\prod_{k\neq i,j} {\mathrm rank}(E_{k}) .
\ee
This example exhibits the general nature of the expansion of the
invariants and the fact that the basis (3-manifold information) is the
same for the Chern-Simons theory and for the \RW model while the
weights (products of charges for the $U(1)$ Chern-Simons theory,
Casimirs for the non-Abelian Chern-Simons theory, integrals of
Chern classes and perhaps other objects in the \RW theory) encode the
differences.

\section{Calculations}\label{calcs}

In this section I will calculate the invariants for 3-manifolds with
$b_{1}(M) \geq 1$. One can do certain calculations for rational
homology spheres, especially for low dimensional $X$, and I
will present some of those here as well. Many of the details of the
calculations are variations on themes taken from \cite{RW}, \cite{T},
\cite{HT} and so I will be somewhat brief here and refer the reader to
the references for more detail.

\subsection{Zero Mode Counting}

Various arguments, (see \cite{RW}, \cite{T} and \cite{HT}) allow one
to conclude that the only relevant parts of the connections that appear in
perturbative calculations of Feynman diagrams from $L_{1}$ (\ref{lag1}),
$L_{2}$ (\ref{lag2}) and (\ref{hwloop}) are
\bea
V_{1} &=& g_{I\oJ}(\f_{0})\, R^{\oJ}_{\; \oK \oL M}(\f_{0})\chi^{I}
\eta^{ \oL}_{0}
d\f^{\oK}_{\perp}\,  \f^{M}_{\perp}
\label{v1} \\
V_{2} &=& -\frac{1}{6}\eps_{IJ}(\f_{0})\, R^{J}_{\; KL\oM}(\f_{0}) \,
\chi^{I}\chi^{K}\chi^{L} \eta^{\oM}_{0}\label{v2}  \\
V_{E} & = & \overline{\partial}_{\overline{J}} \, \omega_{I}(\f_{0}) \,
\chi^{I}\eta^{\overline{J}}_{0}, \label{vE}
\eea
respectively. In
(\ref{vE}) the $0$ subscript means the harmonic part of the field
while the $\perp$ subscript means modes orthogonal to the harmonic
part. To ease the burden of notation, from now on all tensors
are understood to be evaluated on the constant map $\f_{0}$. 

The Feynman diagrams that need to be evaluated now arise from
contractions of all of the possible vertices
\be
\langle V^{r}_{1}V^{s}_{2}V_{E}^{t} \rangle
. \label{vev}
\ee

A constraint comes from the fact that for $\dim_{{\Bbb C}}X =
2n$ the maximal possible product of $\eta^{I}_{0}$ is
$2n$. Consequently,
\be
r+s+t = 2n , \label{select1}
\ee
in order to soak up the $\eta^{\oI}_{0}$ zero modes. In the following
we will look at constraints that arise by counting $\chi^{I}$ zero
modes. The $\chi^{I}$ zero modes can only appear in the vertices
(\ref{v1}-\ref{vE}) with at most one such mode in $V_{1}$, three in
$V_{2}$ and one in $V_{E}$. The number of $\chi^{I}$ zero modes is
$2n\times b_{1}(M)$ (2n because of the holomorphic tangent space label
and $b_{1}(M)$ as it must be a harmonic 1-form on $M$). The largest number
of zero modes that can be soaked up arises when all the $\chi^{I}$
appearing in the vertices are zero modes that is
\be
r + 3s + t = 2n \times b_{1}(M) . \label{select2}
\ee
Together with (\ref{select1}) this implies
\be
s = n \times \left(b_{1}(M) -1 \right) ,
\ee
but is incompatible with (\ref{select1}) if $b_{1}(M) >
3$. Consequently 
\be
Z_{X}^{RW}[M, \prod_{i}{\mathcal O}_{E_{i}}(K_{i})] = 0, \;\;\;
{\mathrm if}\;\; b_{1}(M) > 3. \label{limit}
\ee

It is easy to see that for $b_{1}(M) \geq 1$ that the $\chi^{I}$ that
appears in (\ref{v1}) and (\ref{vE}) must be a zero-mode. In the
following I will take this for granted.

\subsection{Proof of Claim: \ref{cl1}}

Since, in any case $Z_{X}^{RW}[M]=0$ if $b_{1}(M) >3$, (\ref{limit})
partially establishes the claim. When $b_{1}(M)=3$, one finds from the
discussion above that $s=2n$ and $r=t=0$ which means that no vertices
from the link observables can participate in the calculation of the
expectation value of the link observable. So we have
established the claim for $b_{1}(M)=3$. 

If $b_{1}(M)=2$, the rule (\ref{select2}) does not hold (since one
cannot have all three $\chi^{I}$ being harmonic), rather, one can have
at most two $\chi^{I}$ harmonic in $V_{2}$ hence,
\be
r + ps + t = 4n, 
\ee
where $p=0$, $1$ or $2$. $p=0$ and $1$ are ruled out by
(\ref{select1}) leaving only $p=2$, $s=2n$ and $r=t=0$. Once more the
vertices $V_{E}$ do not make an appearance, and so we have established
claim \ref{cl1}.

\subsection{Proof of Claim: \ref{cl2}}

For $b_{1}(M)=1$ the counting of $\chi^{I}$ zero modes tells us that
indeed one of the $\chi^{I}$ appearing in $V_{2}$ must be
a zero mode $\chi^{I}_{0}$. Such a mode is actually decomposable as
\be
\chi^{I}_{0} = c^{I} \, \o
\ee
where $c^{I}$ is an anti-commuting scalar (on $M$) and $\o $ is the
generator of ${\mathrm H}^{1}(M, {\Bbb Z})$. Which means that for a
given knot $K_{i}$ and associated holomorphic vector bundle $E_{i}$ that
\bea
{\mathcal O}_{E_{i}}(K_{i}) &=& {\mathrm Tr}_{V_{i}}\, {\mathrm P}\,
\exp{\left( c^{I}A_{I i} \oint_{K{i}} \o \right)} \nonumber \\
& = & {\mathrm Tr}_{V_{i}}\, \exp{\left( c^{I}A_{I i}\,
\o(K_{i})\right)}. \label{simp}
\eea
The second equality in (\ref{simp}) comes about as follows. Since the
matrix $c^{I}A_{Ii}$ is position independent one can drop the path
ordering. Without the path ordering the integral in the exponent is
really $\oint_{K_{i}}\o$ which is the linking number between the knot
$K_{i}$ and the fundamental cycle Poincare dual to $\o$.

The path integral is still to be performed. However, a glance at
(\ref{simp}) tells us that the insertion of these observables only
effects the zero mode integration of the path integral. The
integration over the other modes has been performed in some generality
in \cite{HT} the result being equation (8.34) in that reference. One
now needs to multiply that result with products of (\ref{simp}) and
integrate over the zero modes of the theory. The integration over the
zero modes turns the objects that appear into differential forms (the
emergence of factors of $2\pi$ is explained in \cite{HT}). After these
gymnastics one obtains claim \ref{cl2}.

\subsection{Proof of Claim: \ref{cl3}}

Here we are interested in $b_{1}(M)=0$ and $n=1$. From the selection
rule (\ref{select1}) we see that
\be
r+s+t = 2 .
\ee
Let us write the final result as a sum of three terms. The first is,
$t=0$, the second $t=1$ and the third comes from $t=2$. 

When $t=0$, we have $r+s=2$ and the only vertices that appear are
those in the calculation of $Z_{X}^{RW}[M]$, so from these diagrams we
get
\be
\left(\prod_{i} {\mathrm rank}E_{i}\right) Z_{X}^{RW}[M] .
\ee

When $t=1$, $r=1$ and $s=0$ or $r=0$ and $s=1$. In the first case the
vertex $V_{1}$ is contracted with itself along the $\f^{i}$ legs. But
this vanishes because the $\f^{i}$ propagator contains a $g^{I\oJ}$
which when contracted with the vertex yields $R^{\oK}_{\; \oJ \oL I}
g^{I\oJ} =0$ since $X$ is Ricci flat. In the second case the vertex
$V_{2}$ is contracted with itself along the two of the three
$\chi^{I}$ legs. This vanishes as well and for the same reason as for
the $V_{1}$ vertex. So there is no contribution from the $t=1$
diagrams. 

For $t=2$ we necessarily have $r=s=0$. This means that we may as well
set the curvature term in the \RW theory to zero for the purposes of
the present calculation. But then the calculation is the same as that
for the $4n$-torus of the previous section. In fact the answer, for
$n=1$ is given in (\ref{rw1}), thus completing the proof of the claim.

\appendix
\section{Coupling to Supersymmetric Quantum Mechanics}

In this appendix I would like to mention one small generalization that
can be made with regards knot invariants. Witten \cite{W} suggested
that the correct way to treat knot observables, Wilson loops, in
Chern-Simons theory is by making use of the Borel-Weil-Bott theorem to
replace Wilson lines by functional integrals over maps from $S^{1}$
into $G/T$. In the present setting a functional integral formulation
of the knot observables is also available. This path integral
representation has a number of uses.

Let $E$ be a holomorphic vector bundle over $X$. One can add to the
\RW action the following supersymmetric action
\be
\oint \, \overline{C}\left( \frac{d}{d t} + d\f^{I}\, \o_{I} - F_{I
\oJ} \chi^{I} \, \eta^{\oJ} \right)\, C \label{sqm}
\ee
where $C$ and $\overline{C}$ are Grassmann odd maps from the knot $K$
to sections of $E$ and $\overline{E}$ respectively. This action is
also $\oQ$ invariant if we set
\be
\oQ \, C = 0 = \oQ \, \overline{C} .
\ee
If we would like this also to exhibit $Q$ invariance then we must take
$E$ to be hyper-holomorphic. If $E$ is hyper-holomorphic then, since
$Q$ acts by a gauge transformation (\ref{gt}), invariance of (\ref{sqm}) is
guaranteed if we perform a gauge transformation on $C$ and
$\overline{C}$, that is,
\bea
Q\, C & = & -\eta^{\oI}\, T^{\, J}_{\oI}\, \o_{J} \, C , \nonumber \\
Q\, \overline{C} & = & \overline{C}\,  \eta^{\oI}\, T^{\, J}_{\oI}\,
\o_{J} .
\eea

One picks out the path ordered exponential by projecting onto the one
particle sector of the theory. This is the equivalent of (\ref{trick})
and can be achieved by placing a projection operator in the
path integral over $C$ and $\overline{C}$. But one is not restricted
to this, rather, one is free to look at any sector of the Hilbert space
that one likes. Consequently, there are many more objects that one can
associate to a knot (and hence to a link). 

Note also that on the 3-manifold $S^{2} \times S^{1}$ one can
essentially squeeze away the non-harmonic modes to be left with a
theory on $S^{1}$ \cite{T}. If one picks the knot $K$ to be $\{x\} \times
S^{1}$ for some, immaterial, point $\{ x\} \in S^{2}$ then the
combined theory, (\ref{rwact}) together with (\ref{sqm}), is a standard
supersymmetric quantum mechanics which
represents the index of the Dolbeault operator coupled to a
holomorphic bundle \cite{AG}.

It would be interesting to have a topological field theory whose
bosonic field is a section of $TX \otimes E$ and not just to couple
$E$ to a knot.

\rnc{\Large}{\normalsize}

\end{document}